\documentclass[12pt,onecolumn]{article}

\usepackage{amsmath}
\usepackage{amssymb}
\usepackage{multirow}
\usepackage{dcolumn}

\usepackage{graphicx}

\usepackage{cite}

\usepackage{color} 

\usepackage{setspace}

\usepackage[labelfont=bf,labelsep=period,justification=raggedright]{caption}

\makeatletter
\renewcommand{\@biblabel}[1]{\quad#1.}
\makeatother

\date{}

\begin{document}

% Title must be 150 characters or less
\begin{flushleft}
{\Large
\textbf{Can Google searches help nowcast and forecast unemployment rates in the Visegrad Group countries?}
}
% Insert Author names, affiliations and corresponding author email.
\\
Jaroslav Pavlicek $^{1}$ and Ladislav Kristoufek$^{1,2,3,\ast}$
\\
\bf{1} Institute of Information Theory and Automation, Academy of Sciences of the Czech Republic, Pod Vodarenskou vezi 4, Prague 8, 182 08, Czech Republic, EU
\\
\bf{2} Institute of Economic Studies, Charles University, Opletalova 26, 110 00, Prague, Czech Republic, EU
\\
\bf{3} Warwick Business School, University of Warwick, Coventry, West Midlands, CV4 7AL, United Kingdom, EU
\\\
$\ast$ E-mail: kristouf@utia.cas.cz
\end{flushleft}

\doublespacing

% Please keep the abstract between 250 and 300 words
\section*{Abstract}
Online activity of the Internet users has been repeatedly shown to provide a rich information set for various research fields. We focus on the job-related searches on Google and their possible usefulness in the region of the Visegrad Group -- the Czech Republic, Hungary, Poland and Slovakia. Even for rather small economies, the online searches of their inhabitants can be successfully utilized for macroeconomic predictions. Specifically, we study the unemployment rates and their interconnection to the job-related searches. We show that the Google searches strongly enhance both nowcasting and forecasting models of the unemployment rates.

\section*{Introduction}

Online activity has become an inherent part of the modern society and the way of living among its members. The Internet provides a vast amount of information to its users as well as an aid and assistance in times of need. During the current financial and following economic and production crises, most of the developed as well as developing economies have been hit by an economic downturn which is usually tightly connected with a growing unemployment. Job loss can be a very traumatizing experience with long lasting impact on one's life. Seeking a new job then becomes an integral part of an everyday life. In the current digitalized era, the job seeking does not restrict itself to job offices but the seekers (as well as potential employers) more frequently turn to the Internet as a source of information and new possibilities. As such, the job seekers leave a digital track of their activity.

Analysis and examination of various patterns of the online activity have become a fruitful branch of research in the last years with some exciting applications such as elections \cite{Metaxas2012}, investment allocation \cite{mondria2010,kristoufek2013a}, private consumption \cite{Vosen2011} and consumers' behavior \cite{Goel2010}, future orientation \cite{Preis2012a}, earnings announcements \cite{drake2012}, diseases spreading \cite{polgreen2008,ginsberg2009,Carneiro2009,Seifter2010,Dugas2012}, and economics and finance \cite{Preis2010,varian2012,Bordino2012,kristoufek2013b,preis2013,Moat2013,Curme2014}. Turning back to the unemployment and its possible examination utilizing the online activity of the Internet users, there has been some research done in the area as well focusing primarily on the Google engine search queries. The first study focusing on the possible connection between Google searching activity and unemployment rates examining the series in Germany shows usefulness of adding search queries data into the models \cite{askitas2009}. Following research \cite{varian2009b,bughin2011,varian2013} analyzes connection between the queries and claims for unemployment benefits in the USA and the unemployment rate itself has been studies as well \cite{damuri2010,damuri2012}. Even job search activity index based on the Google search data has been developed \cite{baker2011}. Most of these studies focus on the US economy and its modeling while the other economies are studied rather marginally \cite{chadwick2012,fondeur2012}.

Here, we focus on possible connection between job-related search queries on the Google search engine and the unemployment rate in countries of the so-called Visegrad Group (the Czech Republic, Hungary, Poland and Slovakia). Our contributions lay in the following. First, we focus on a set of countries which would be normally treated as a marginal one and thus not much studied. However, if the utility of the online search activity (and specifically the Google searching) is to be claimed, its efficiency should be shown not only on developed and well covered countries but also on the smaller ones and the results might prove useful to all policy makers even in such regions. Second, we provide a careful and step-by-step procedure to the unemployment modeling focusing not only on simple correlations but also nowcasting, forecasting and causality. And third, a cross-countries comparison is delivered which is rather unique in comparable studies focusing primarily on one specific country.

% Results and Discussion can be combined.
\section*{Results}

\begin{figure}[htbp]
\center
\begin{tabular}{c}
\includegraphics[width=4.5in]{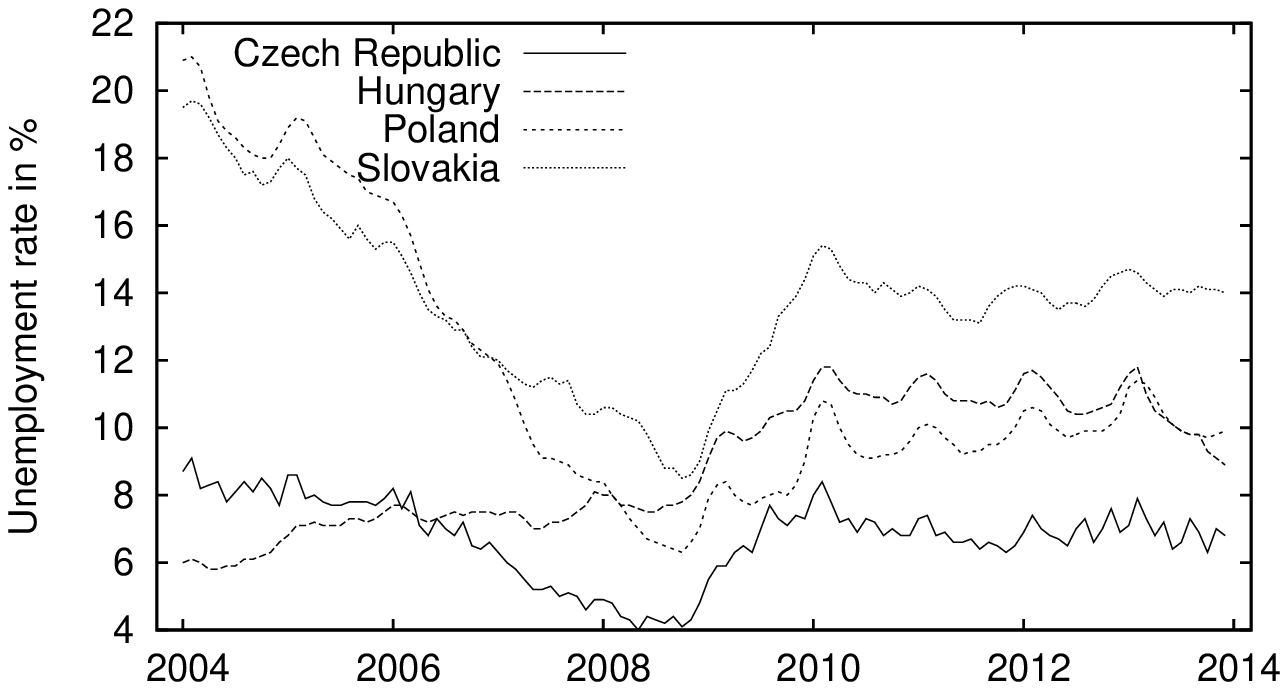}\\
\end{tabular}
\caption{\footnotesize\textbf{Unemployment rate in the Visegrad countries.} The group of countries is evidently quite heterogenous in the unemployment rates. The Hungarian rate starts at the lowest level but increases stably during the whole period. The Czech rate begins at quite low levels and decreases up to the outbreak of the financial crisis when the rate surges up until 2010 after which it remains quite stable. The Polish and Slovakian rates commence at very high levels of unemployment  which go down again up until the outbreak of the crisis after which they change the trends similarly to the Czech rate.\label{fig_U}}
\end{figure}

The unemployment rates have undergone quite heterogenous evolution in the analyzed countries (Fig. \ref{fig_U}). In the Czech Republic, the rate ranged between 4\% and 9\% between years 2004 and 2013. Initially, there was a significant downward trend from year 2004 to 2008 when the rate dropped from 9\% to 4\%. As the recession hit the Czech Republic in 2008, the rate started to rise to reach its new maximum of 8.5\% in 2010. Since then, the unemployment rate fluctuated between 7\% and 8.5\%. The Hungarian unemployment rate was steadily rising from the year 2004 to 2010 where it reached its new maximum of nearly 12\%. After that the rate fluctuated for almost 3 years between 10.5\% and 12\% to start declining in the year 2013. The unemployment in Poland experienced a steady decline from the astronomical rate of nearly 22\% in the year 2004 to 6\% in 2009. However, as the recession hit Poland, the unemployment rate began rising again. With some minor fluctuations, it smoothly increased to the current level of approximately 10\%. And in Slovakia, the unemployment rate seems to have a similar pattern as the one of the Czech Republic, although on a different scale. In 2004, Slovakia had an unemployment rate of almost 20\%. This rate linearly decreased to 8\% in 2009. With the hit of recession, the unemployment rate quickly escalated to 16\% around which it has been fluctuating until today.

\begin{figure}[htbp]
\center
\begin{tabular}{c}
\includegraphics[width=4.5in]{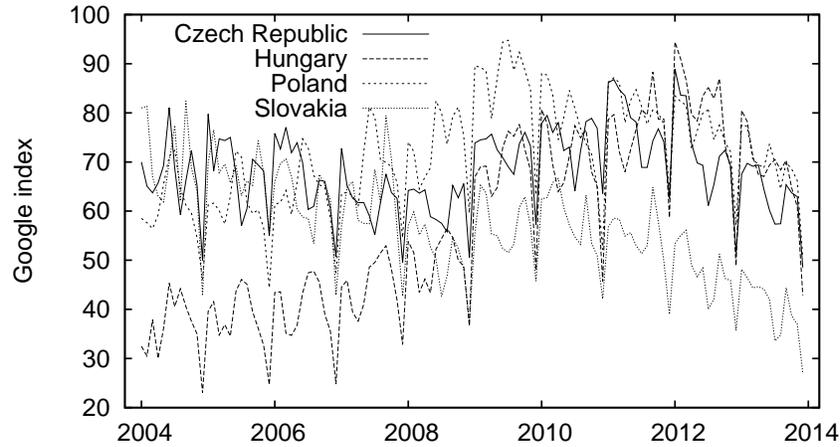}\\
\end{tabular}
\caption{\footnotesize\textbf{Google search queries for the job-related terms in the Visegrad countries.} The patterns are again quite heterogenous and the connection between the Google searches and the unemployment rates can be observed for the Czech and Hungarian rates. For the other two, the connection is not visible by the naked eye. Detailed treatment of the interconnections is given in the Results section of the text.\label{fig_Google}}
\end{figure}

The evolution of the Google searches is illustrated in Fig. \ref{fig_Google}. There are evident seasonal patterns in all four series. Hungary is characterized by quite regularly increasing trend in the Google searches whereas Slovakia shows the opposite and the remaining two analyzed series remain quite stable in time. Even though there seems to be some connection between the Google searches and the unemployment rates for the Czech Republic and Hungary visible by the naked eye, we can hardly claim any relationship without a proper analysis.

\subsection*{Basic relationship}

As the initial step, we present the results of the stationarity tests which tell us whether we should analyze the original series or some of their transformations. In Tab. \ref{table:stat}, we show the results of the ADF and KPSS tests (see the Methods section for more details) for the original as well as the logarithmic series and their first differences. The outcome is quite straightforward as we do not reject unit roots for either of the original series (or their logarithmic transformation for the Google searches, we do not examine the logarithmic transformation for the unemployment time series as these are already in the percentage representation). Further testing, which is not reported here, shows no cointegration relationship between the unemployment and the search queries series so that we need to proceed with the first differences of the series. For most of the cases, we support stationarity of the first differences. In the analysis, we further proceed with the first differences of the unemployment rate and the first logarithmic differences of the Google searches. We opt for this combination as the pair of percentage representation and logarithmic transformation allows for a straightforward interpretation as an elasticity, i.e. as a proportional relationship.

For the very basic relationship between the unemployment rate and the intensity of the job-related searches on Google, we study the following equation
\begin{equation}
\label{eq:basic}
\begin{aligned}
\Delta \text{UR}_t &= \alpha_0 +  \alpha_1 \Delta \log(\text{GI})_t + \varepsilon_{t}
\end{aligned}
\end{equation}
where $\Delta \text{UR}_t$ and $\Delta \log(\text{GI})_t$ stand for the first difference of an unemployment rate at time $t$ and the first logarithmic difference of the Google searches at time $t$, respectively, for a given country, and $\varepsilon_t$ is an error term.

The elasticity between the Google searches and unemployment rate from Eq. \ref{eq:basic} is estimated at 0.5538 (with the $p$=value of 0.0533), 0.2056 (0.0726), 0.3317 (0.2163) and 0.4630 (0.0062) for the Czech Republic, Hungary, Poland and Slovakia, respectively, with the heteroskedasticity and autocorrelation consistent (HAC) standard errors. The proportional relationship thus varies across the analyzed countries but it remains positive for all four and statistically significant for three out of four (at least at the 10\% significance level). Specifically, the relationship is very strong for the Czech Republic and Slovakia with the value around 0.5. This shows that the changes in the unemployment rate are well projected into the online search queries for the vacancies and job-related terms. Studying the connection between these two variables thus seems promising and worth further utilization and investigation.

\subsection*{Nowcasting}

Macroeconomic time series, such as the unemployment rates, have a special property which is not present for financial series or other series in natural sciences -- they are available with a pronounced lag. This is due to the data processing and collection which usually take several months and even after such period, there are sometimes corrections to the reported values. Such characteristic makes a series, which is available immediately without any lag and which is strongly correlated with the variable of interest, very useful for forecasting the present value of the variable without waiting for several months. Such forecasting the present is usually referred to as ``nowcasting''.

In the previous section, we have shown that the Google searches for job-related terms are significantly correlated with the unemployment rate which makes the search queries potentially useful for nowcasting of the unemployment. As a nowcasting model, we consider the following one

\begin{equation}
\label{eq:ARnow}
\begin{aligned}
\Delta \text{UR}_t &= \beta_0 + \sum_{i=3}^{12}{\beta_{i} \Delta \text{UR}_{t-i}}  + \sum_{j=0}^{12}{\gamma_{j} \Delta \log(\text{GI})_{t-j}} + \varepsilon_{t}
\end{aligned}
\end{equation}
where the unemployment rate is assumed to be available with a three months lag. We again consider the differenced series due to stationarity issues discussed above. Both series are kept to the lag of 12 months which controls for the seasonal pattern in both the series.

The results of the nowcasting models are summarized in Tab. \ref{table:now}. There we show the adjusted $R^2$ ($\bar{R}^2$) as a measure of the models' quality controlling for the number of explanatory variables. We observe that for all countries, the inclusion of the Google series enhances the model strongly. The $\bar{R}^2$ increases by approximately a third for all countries but Poland for which it increases slightly less. Nonetheless, inclusion of the search queries improves the model for all countries significantly as is reported by the $F$-statistics for the insignificance of the searches. All series are jointly significant even at the 1\% level.

\subsection*{Forecasting \& Causality}

The nowcasting results are very promising and they illustrate usefulness of the Google searches series. However, we are also interested whether such usefulness is mainly due to the unavailability of the unemployment data or whether the search queries data provide additional informative value as well. To do so, we also undergo a standard forecasting exercise where we practically hypothesize what would happen were the unemployment data available straightaway. If the Google series improve even such hypothetical model, we conclude that the search queries data bring additional information to the model in addition to being strongly correlated with the changes in the unemployment rate by itself.

For the forecasting exercise, we utilize the standard vector autoregressive model (VAR, see the Methods section for more details). The specific model takes the following form
\begin{equation}
\label{eq:VAR}
\begin{aligned}
\Delta \text{UR}_t &= \beta_{01} + \sum_{i=1}^{12}{\beta_{1i} \Delta \text{UR}_{t-i}}  + \sum_{i=1}^{12}{\gamma_{1i} \Delta \log(\text{GI})_{t-i}} + \varepsilon_{1t} \\
\Delta \log(\text{GI})_t &= \beta_{02} + \sum_{i=1}^{12}{\beta_{2i} \Delta \text{UR}_{t-i}}  + \sum_{i=1}^{12}{\gamma_{2i} \Delta \log(\text{GI})_{t-i}} + \varepsilon_{2t}
\end{aligned}
\end{equation}
and it is compared to a simple autoregressive model of unemployment
\begin{equation}
\label{eq:AR}
\begin{aligned}
\Delta \text{UR}_t &= \delta_{0} + \sum_{i=1}^{12}{\delta_{1i} \Delta \text{UR}_{t-i}} + \nu_{t}.
\end{aligned}
\end{equation}

For the comparison purposes, we use two measures of the forecasting quality -- root mean squared error and mean absolute error (RMSE and MAE, respectively, see the Methods section for more details). These measures are very straightforward -- the lower they are the better performing the model is. In addition, we utilize the Diebold-Mariano test \cite{diebold1995} which compares the forecasting performance of two models with the null hypothesis of the models performing the same (see the Methods section for more details). The model is estimated on the series between January 2004 and December 2012 and the forecasting period is set between January and December 2013.

The summary of the forecasting performances is given in Tab. \ref{table:fore}. There we can see that for all countries, the forecasting performance of the models increases strongly with the addition of the Google searches. This is further supported by the results of the Diebold-Mariano test which gives significant results, i.e. the model using the Google data outperforms the ones without them, for all countries at at least the 5\% significance level. The online search data thus evidently provide an additional informative value to the unemployment modeling.

As the last step of the analysis, we provide a causality examination. We are thus interested in the specific relationship between the two analyzed series. Concretely, we examine whether the increasing unemployment causes people to look up the job-related terms more, or the increased online activity signalizes potential tensions on the job market, or both ways, or none. To do so, we utilize the Granger causality framework (see the Methods section for more details) which is built on the VAR analysis. The results are summarized in Tab. \ref{table:fore}. Note that the null hypothesis of the Granger causality is ``no Granger causality''. Therefore, if the null hypothesis is rejected, the causality is claimed to be found. The findings are quite homogenous. For three out of four countries (Hungary being the exception), we report causality in both directions. The influence thus goes from both directions and the series strongly influence each other.

\section*{Discussion}

Online activity of the Internet users has been proven useful in various fields. Nowcasting the unemployment rate is one of these fields. Contrary to the prevailing trend in the literature focusing on the well-developed (Western) countries, we have focused on utilizing the job-related Google searches in the Visegrad Group countries, i.e. the Czech Republic, Hungary, Poland and Slovakia. Even though the data availability and utilization of the Internet might not be as widespread in the region as one would expect for the developed countries, we have shown that in fact the online searches provide a very strong basis for the unemployment modeling.

In summary, we have shown that the basic dynamics of the Google searches for the job-related terms closely follows the one of the unemployment rates. Further, we have utilized this idea to successfully nowcast the unemployment rates using the current and lagged values of the Google searches. Such results have been shown to be caused not simply by the fact that the unemployment rates are not immediately available but also by the additional informative value of the online searches. Our findings indicate that the information left online by the Internet users can be easily utilized even for small or medium countries such as the ones of the Visegrad Group.

\section*{Methods}

\subsection*{Data}

The monthly unemployment data for the Czech Republic, Hungary, Poland and Slovakia have been obtained from the Eurostat database. The basis of unemployment measurement among the EU countries lies in the EU Labour Force Survey (EU LFS) -- a continuous and harmonized household survey, which is in accordance with the EU legislation carried out in each member state. The monthly data from Eurostat are estimates based on the results of EU LFS. Since there are no legal obligations for the EU countries to deliver monthly data, these data are often interpolated/extrapolated using national survey or registered unemployment data.

According to Eurostat, an unemployed person is defined as someone aged between 15 and 74 without work during the reference week who is available to start working within two weeks and who has actively sought employment at some time during the last four weeks. In our analysis, we use the general (both sex, 15-74 years old) raw (not seasonally adjusted) unemployment rate. We do this since we do not know the method used for the seasonal adjustment and the Google data are not seasonally adjusted either.

The Google search queries data have been downloaded from the Google Trends webpage. As languages of the studied countries differ, we have looked for various terms. As the Czech, Polish and Slovakian are all Slavonic languages, the searched words are very similar or even the same. For Czech, we searched for ``pr\'{a}ce'', for Polish ``praca'' and for Slovakian ``pr\'{a}ce'' as well. For Hungarian, we used term ``\'{a}ll\'{a}s''. For the Slavonic languages, the terms are equivalent to ``job'' or  ``work'', and for Hungarian, it is close to ``job'' or  ``work'' but rather in a sense of looking for it. The term ``\'{a}ll\'{a}s'' provides better results than a more straightforward ``munka'' which would be closer to a more standard meanings of terms ``job'' or  ``work'.

The weekly series obtained from the Google Trends site have been transformed to the monthly series on a basis of the number of days in the month basis. All series, both of the unemployment rate and the Google searches, are studied between January 2004 and December 2013.

\subsection*{Stationarity}
Stochastic process $\{z_t\}$ is stationary if for every collection of time indices $1 \leq t_1 < t_2 < t_m$, the joint probability distribution of ($x_{t_1}$, $x_{t_2}$, ..., $x_{t_m}$) is the same as the joint probability distribution of ($x_{t_{1+h}}$, $x_{t_{2+h}}$, ..., $x_{t_{m+h}}$) for all integers $h\geq 1$ \cite{wooldridge2008}. To test for stationarity, we utilize the Augmented Dickey-Fuller (ADF) test \cite{dickey1979} and the KPSS test \cite{kwiatkowski1992}. The tests have opposite null hypotheses so that they provide a complementary pair which is commonly used for stationarity testing.

In the ADF procedure \cite{dickey1979}, the OLS regression is run on
\begin{equation}
\label{eq:adf}
\Delta z_t = \alpha_0 + \theta z_{t-1} + \gamma t + \Delta z_{t-1} + \Delta z_{t-2} + \cdots + \Delta z_{t-p} + \varepsilon_t
\end{equation}
in order to perform the test, where $\alpha_0$ and  $\gamma t$ are an intercept and a time trend, respectively, and $p$ represents the lag order.
The null hypothesis under which the series contains a unit root is found for
$$H_0: \theta = 0$$
against the alternative
$$H_A: \theta < 0.$$
The ADF test statistics is then computed as usual $t$-statistics, which, however, follows a more complicated distribution under the null hypothesis. Due to the relative short time series, we set the number of lags arbitrarily to three.

The null hypothesis of the KPSS test \cite{kwiatkowski1992} is opposite to the one of the {ADF} test, i.e. the KPSS test has the null hypothesis of stationarity. The test is based on the {OLS} regression of the series $\{z_t\}$
\begin{equation}
\label{eq:kpss}
z_t = \alpha_0 + \gamma t + k\sum_{i=0}^{t}{\xi_i} + \varepsilon_t
\end{equation}\\
where $\alpha_0$ and  $\gamma t$ again represent an intercept and a time trend, respectively, and $\xi_i$ are independent and identically distributed random variables with a zero mean and a unit variance. The null hypothesis of stationarity is found for
$$H_0: k = 0$$
against the alternative
$$H_A: k \neq 0.$$
The KPSS test statistic is defined as
$$KPSS = \frac{\sum_{t=1}^{n}{S_t^2}}{n^2\hat{\omega}_T^2}$$
where $S_t$ is partial sum of residuals
$$S_t = \sum_{i=1}^{t}{\hat{\varepsilon}_i}$$
and $\hat{\omega}_T^2$ is an estimator of the spectral density at a frequency zero.

\subsection*{Vector autoregression}
Vector autoregression (VAR) is simply a system of temporally dependent series. More precisely, denote the number of variables $k$ and the length of the series $T$, then VAR of order $p$ is generally represented by equation
\begin{equation}
\label{eq:var}
y_t = \alpha + A_1 y_{t-1} + A_2 y_{t-2} + \cdots + A_p y_{t-p} + \varepsilon_t
\end{equation}
where $y_t$ and $\varepsilon_t$ are $k \times T$ matrices representing the studied series and residuals, respectively, $\alpha$ represents a vector of constants and $A_i$ are time invariant matrices replacing the traditional $\beta_i$ coefficients.
The selection of appropriate lag order $p$ is usually based on a specific information criterion. 

In the VAR framework, the Granger causality concept is usually used as well. The causality testing simply stems in testing the joint significance of one of the variables in the equation for some other variable. The testing procedure is thus an $F$-test for joint significance of a specific variable. In needs to be noted that such causality is strictly statistical and it should be always treated with caution.

\subsection*{Forecasting}
To compare forecasting accuracy of the proposed models, we utilize three measures -- mean absolute error (MAE), root mean squared error (RMSE) and the Diebold-Mariano test \cite{diebold1995}.

MAE measures the average value of absolute losses. In other words, it gives an average deviation of forecast from realized value in absolute terms. It is given by the equation 
\begin{equation}
\label{eq:mae}
MAE = \frac{1}{T} \sum_{i=1}^{T} |f_i - y_i| = \frac{1}{T} \sum_{i=1}^{T} a_i
\end{equation}
where $f_i$ stands for the predicted value, $y_i$ is the actual value and $a_i = |f_i - y_i|$.

RMSE is quite similar to the mean absolute error as it is simply a square root of the mean squared error, and it is defined as
\begin{equation}
\label{eq:rmse}
RMSE = \sqrt{\frac{1}{T} \sum_{i=1}^{T} (f_i - y_i)^2} = \sqrt{\frac{1}{T} \sum_{i=1}^{T} s_i} 
\end{equation}
where $f_i$ stands for the predicted value $y_i$ is the actual value and $s_i = (f_i - y_i)^2$.

Diebold \& Mariano \cite{diebold1995} propose a test to compare the predictive accuracy of two competing forecasts. Let $\{\varepsilon^1_t\}_{t_0}^T$ and $\{\varepsilon^2_t\}_{t_0}^T$ be the sequences of forecast errors losses from two competing forecasting measures by particular loss function (e.g. absolute error loss as $a_i$ in Eq. \ref{eq:mae} or squared error loss as $s_i$ in Eq. \ref{eq:rmse}). The null and alternative hypotheses are then stated as
$$H_0: \mathbb{E}\{\varepsilon^1_t\}_{t_0}^T = \mathbb{E}\{\varepsilon^2_t\}_{t_0}^T$$
$$H_A: \mathbb{E}\{\varepsilon^1_t\}_{t_0}^T \neq \mathbb{E}\{\varepsilon^2_t\}_{t_0}^T.$$
The Diebold-Mariano test assesses the accuracy based on the loss differential
$$d_t = \{\varepsilon^1_t\}_{t_0}^T - \{\varepsilon^2_t\}_{t_0}^T$$
and the underlining null
$$H_0: \mathbb{E}\{d_t\} = 0.$$
The Diebold-Mariano statistics is then
$$S = \frac{\bar{d}}{\sqrt{\widehat{LRV}_{\bar{d}}/T}}$$
where $\bar{d}$ is the mean loss differential
$$LRV_{\bar{d}} = \gamma_0 + 2 \sum_{j=1}^{\infty}{\gamma_j}, \;\; \gamma_j = \text{cov}(d_t, d_{t-j})$$
and $\widehat{LRV}_{\bar{d}}$ is a consistent estimate of the asymptotic (long-run) variance of $\sqrt{T}\bar{d}$. Under the null hypothesis, the testing statistic goes to a standard normal distribution so that $S \stackrel{A}{\sim} N(0,1)$ \cite{diebold1995}.

% You may title this section "Methods" or "Models". 
% "Models" is not a valid title for PLoS ONE authors. However, PLoS ONE
% authors may use "Analysis" 

%\section*{Materials and Methods}

% Do NOT remove this, even if you are not including acknowledgments
\section*{Acknowledgments}

The research leading to these results has received funding from the European Union's Seventh Framework Programme (FP7/2007-2013) under grant agreement No. FP7-SSH-612955 (FinMaP) and the Czech Science Foundation project No. P402/12/G097 ``DYME -- Dynamic Models in Economics''.

%\section*{References}
% The bibtex filename
\bibliography{PLoS}

\onecolumn

\section*{Figures}

\newpage

\section*{Tables}

\begin{table}[!htbp]
\begin{center}
\caption{Stationarity testing}
%\vspace{0.5cm}
\label{table:stat}
\begin{tabular}{l |cccc}
\hline\hline 
&Czech Rep. & Hungary & Poland & Slovakia\\
\hline
\multicolumn{5}{c}{\textit{ADF test}}\\
\hline
Unemployment&-1.6066&-1.78611&-2.6739$^{\ast}$&-2.6438$^{\ast}$\\
- first difference&-5.3860$^{\ast\ast\ast}$&-4.5134$^{\ast\ast\ast}$&-3.5267$^{\ast\ast\ast}$&-4.2349$^{\ast\ast\ast}$\\
Google&-2.6399$^{\ast}$&-1.76434&-2.1745&-1.4327\\
- logarithm&-2.6504$^{\ast}$&-2.0239&-2.3280&-0.9082\\
- difference&-11.2221$^{\ast\ast\ast}$&-10.2869$^{\ast\ast\ast}$&-11.1560$^{\ast\ast\ast}$&-10.5487$^{\ast\ast\ast}$\\
- logarithmic difference&-11.3131$^{\ast\ast\ast}$&-10.7993$^{\ast\ast\ast}$&-11.0750$^{\ast\ast\ast}$&-10.5391$^{\ast\ast\ast}$\\
\hline
\multicolumn{5}{c}{\textit{KPSS test}}\\
\hline
Unemployment&0.5399$^{\ast\ast}$&2.5995$^{\ast\ast\ast}$&1.7946$^{\ast\ast\ast}$&0.7507$^{\ast\ast\ast}$\\
- first difference&0.1932&0.2848&0.6708$^{\ast\ast}$&0.5673$^{\ast\ast}$\\
Google&0.4208$^{\ast\ast}$&2.5994$^{\ast\ast\ast}$&1.3281$^{\ast\ast\ast}$&2.2879$^{\ast\ast\ast}$\\
- logarithm&0.4048$^{\ast}$&2.6349$^{\ast\ast\ast}$&1.3596$^{\ast\ast\ast}$&2.2123$^{\ast\ast\ast}$\\
- difference&0.0730&0.1406&0.1358&0.0436\\
- logarithmic difference&0.0818&0.1201&0.1362&0.0821\\
\hline    \hline
\end{tabular}
%\vspace{-0.7cm}
\end{center}
\end{table}

\begin{table}[!htbp]
\begin{center}
\caption{Nowcasting summary}
%\vspace{0.5cm}
\label{table:now}
\begin{tabular}{c|c|cccc}\hline\hline 
&& Czech Rep. & Hungary & Poland & Slovakia\\
 \hline
\multirow{2}{*}{$\bar{R}^2$}&without Google&0.2796&0.3590&0.4673&0.1605\\
&with Google&0.3763&0.4469&0.5521&0.2193\\
\hline
\multirow{2}{*}{Google insignificant}&$F$-stat&5.8507&2.5828&2.5251&2.7135\\
&$p$-value&0.0000&0.0089&0.0057&0.0031\\
\hline    \hline
\end{tabular}
%\vspace{-0.7cm}
\end{center}
\end{table}

\begin{table}[!htbp]
\begin{center}
\caption{Forecasting and causality summary}
%\vspace{0.5cm}
\label{table:fore}
\begin{tabular}{c|c|cccc}\hline\hline 
&& Czech Rep. & Hungary & Poland & Slovakia\\
 \hline
\multirow{3}{*}{RMSE}&no Google&0.3686&0.3156&0.1990&0.1907\\
&Google&0.3216&0.2742&0.1467&0.1359\\
&change&-12.75\%&-13.11\%&-26.28\%&-28.73\%\\
 \hline
\multirow{3}{*}{MAE}&no Google&0.3100&0.2183&0.1437&0.1638\\
&Google&0.2744&0.1889&0.1164&0.1009\\
 &change&-11.50\%&-13.49\%&-19.01\%&-38.41\%\\
  \hline
\multirow{2}{*}{DM test}&test statistic&2.4160&2.8870&2.0220&1.7750\\
&$p$-value&0.0078&0.0019&0.0216&0.0379\\
   \hline    \hline
\multirow{2}{*}{Google $\rightarrow$ Unemployment}&test statistic&6.0311&1.1163&3.8285&3.1555\\
&$p$-value&0.0000&0.3614&0.0002&0.0012\\
    \hline
\multirow{2}{*}{Unemployment $\rightarrow$ Google}&test statistic&2.5545&1.4951&2.5635&2.5641\\
&$p$-value&0.0073&0.1469&0.0071&0.0071\\
\hline    \hline
\end{tabular}
%\vspace{-0.7cm}
\end{center}
\end{table}

%\begin{table}[!ht]
%\caption{
%\bf{Table title}}
%\begin{tabular}{|c|c|c|}
%table information
%\end{tabular}
%\begin{flushleft}Table caption
%\end{flushleft}
%\label{tab:label}
% \end{table}

\end{document}